\documentclass[prd,superscriptaddress,nofootinbib,showkeys,12pt]{revtex4-1}

\usepackage[utf8]{inputenc}
\usepackage[T1]{fontenc}
\usepackage[english]{babel}
\usepackage{hyperref}

\usepackage{amsmath}
\usepackage{amssymb}
\usepackage{amsfonts}
\usepackage{mathrsfs}
\usepackage{graphicx}
\usepackage{xcolor}
\usepackage{bm}
\usepackage{esdiff}
\usepackage{commath}
\usepackage{mathtools}
\usepackage{tikz-cd} 
\usepackage[separate-uncertainty]{siunitx}
\usetikzlibrary{decorations.pathmorphing}

\usepackage{comment}

\usepackage{changepage}
\usepackage{mathtools}

\newcommand{\new}[1]{#1}

\newcommand{\Coeff}{C}
\newcommand{\CoeffIndex}[1]{C_{#1}}
\newcommand{\Cdiff}[3]{\Coeff_{#1:{#2}}{}^{#3}}

\newcommand{\Ecoeff}[2]{{E^{#1}}_{#2}}
\newcommand{\Fcoeff}[3]{{{F^{#1}}_{#2}}^{#3}}

\newcommand{\evalphi}{\ensuremath{|_{\varphi^A(a,r,\theta)}}}

\begin{document}
\title{Symmetric gravitational closure}
\author{Maximilian Düll}
\affiliation{Zentrum für Astronomie der Universität Heidelberg, Astronomisches Rechen-Institut, Philosophenweg~12, 69120 Heidelberg}
\author{Nils L. Fischer}
\affiliation{Max-Planck-Institut für Gravitationsphysik, Am Mühlenberg~1, 14476 Golm, Germany}
\author{Björn Malte Schäfer}
\affiliation{Zentrum für Astronomie der Universität Heidelberg, Astronomisches Rechen-Institut, Philosophenweg~12, 69120 Heidelberg}
\author{Frederic P. Schuller}
\affiliation{Department of Applied Mathematics, University of Twente,\\P.O. Box 217, 7500 AE Enschede, The Netherlands}

\begin{abstract}
\new{We show how to exploit symmetry assumptions to determine the dynamical equations for the particular geometry that underpins given matter field equations. 
The procedure builds on the gravitational closure equations for matter models without any a priori assumption of symmetry. 
It suffices to illustrate the symmetrization procedure for a Klein-Gordon field equation on a Lorentzian background, for which one obtains the Friedmann equations, without ever having known Einstein's equations, by careful imposition 
of maximal cosmological symmetry directly on the pertinent gravitational closure equations.
This method of finding the family of symmetry-reduced gravi\-tational field equations that are compatible with given matter dynamics directly generalizes to any Killing symmetry algebra, matter models beyond the standard model and indeed tensorial spacetime geometries beyond Lorentzian metrics.   
}
\end{abstract}


\maketitle
This letter presents a proof-of-concept study for the simplified derivation of gravitational dynamics from given matter field equations, once Killing symmetries of the spacetime geometry are assumed. 

Without symmetry assumptions, the dynamics of a physically constrained class of matter actions restrict the possible dynamics of the underlying geometry so severely, that the dynamics for this geometry can be determined constructively \cite{GravClosure18}. More precisely, any matter action $S_\textrm{\tiny matter}[\Phi,G)$, which is local in some matter field $\Phi$ and ultralocal in a tensor field $G$ and satisfies three algebraic physicality conditions \cite{RSS} constructively determines the causally compatible gravitational actions $S_\textrm{\tiny gravity}[G]$ for the tensor field $G$. Adding the latter action to the former closes the matter dynamics gravitationally, since variation of the total action with respect to $\Phi$ now recovers the stipulated matter field equations while its variation with respect to $G$ yields the gravitational \new{field} equations for the pertinent tensorial geometry $G$ sourced by the very matter field dynamics at play,
\begin{equation}\label{alleqns}
\frac{\delta S_\textrm{\tiny matter}}{\delta \Phi}[\Phi,G) = 0 \qquad \textrm{ and } \qquad \frac{\delta S_\textrm{\tiny gravity}}{\delta G}[G] = - \frac{\delta S_\textrm{\tiny matter}}{\delta G}[\Phi,G)\,.
\end{equation}
\new{The procedure to construct the action $S_\textrm{\tiny gravity}$ from a given action $S_\textrm{\tiny matter}$ is divided into two steps. First, a series of ultimately simple algebraic calculations starting from $S_\textrm{\tiny matter}$ is carried out resulting in the coefficient functions of a countable system of partial differential equations. This system has to be solved in the second step. This solution then} provides the Lagrangian density of the desired gravitational action. 

An instance of a matter action, for which this gravitational closure procedure can be performed manually and with acceptable calculational effort, is \new{the Klein-Gordon action} 
$$S_\textrm{\tiny matter}[\phi,g) = \int_M \mathrm d^4x\,\sqrt{-g} \,\left(\frac{1}{2}\,g^{ab} \,\partial_a\phi \, \partial_b\phi - m^2\phi^2\right)\,,$$ where $\phi$ is a scalar field \new{and $g$ a metric tensor on a four-dimensional manifold $M$, which satisfies the aforementioned three weak physicality conditions for the matter action if and only if the metric has Lorentzian signature.} 
Gravitational closure \new{of these Klein-Gordon dynamics was shown \cite{GravClosure18,GSWW} to yield the} well-known two-parameter family
$S^{\kappa,\Lambda}_\textrm{\tiny gravity}[g] = \kappa \int \mathrm d^4x \,\sqrt{-g} \,(R[g] - 2 \Lambda)$ of Einstein-Hilbert actions, where $R$ is the Ricci curvature scalar of the metric tensor field $g$. At present prohibitively difficult, \new{in contrast}, is the manual calculation of the gravitational closure of linear birefringent electrodynamics \citep{HehlObukhov, SchullerWitteWohlfarth}, which is given by the matter action \new{$S_\textrm{\tiny matter}[A,G) = -\frac{1}{2}\, \int \mathrm d^4x\,  \omega_G \, G^{abcd} \partial_{[a} A_{b]} \partial_{[c} A_{d]}$} for a one-form gauge potential $A$, which employs a fourth rank tensor field $G^{abcd}$ with the symmetries $G^{abcd}=G^{cdab}$ and $G^{abcd}=-G^{bacd}$ and a scalar density $\omega_G$ constructed from it as its background geometry. While the mentioned countable set of gravitational closure equations is set up straightforwardly, see \cite{GravClosure18}, the bottleneck is their actual solution. 

The purpose of the present note is to demonstrate how the typically \new{difficult} 
solution of gravitational closure equations can be significantly simplified by implementing 
spacetime symmetries already when solving this countable set of partial differential closure equations, \new{rather than later at the level of the resulting gravitational action or indeed the gravitational field equations.} This will of course only lead to \new{a} 
symmetry-reduced gravitational action and is thus obviously \new{only} viable exactly under the same known conditions \cite{FelsTorre} of symmetric criticality \cite{Palais} that afford one to impose symmetry assumptions interchangeably either at the level of an action or at the level of the ensuing field equations.

\begin{figure}[h]
\begin{equation*}
\begin{tikzcd}[row sep=huge, column sep=huge]
\begin{array}{c}\textrm{\small Klein-}\\[-9pt] \textrm{\small Gordon}\\[-9pt] \textrm{\small action}\end{array}
\arrow[d,"\textrm{\small set up}"] 
\arrow[dr,dashrightarrow,bend left,"\textrm{\small add}"]
& & & \\ 
\begin{array}{c}\textrm{\small closure}\\[-9pt] \textrm{\small equations}\end{array}
\arrow[r,"\textrm{\small solve}"] 
\arrow[d,"\textrm{\small symmetrize}"]
&
\begin{array}{c}\textrm{\small Hilbert}\\[-9pt] \textrm{\small action}\end{array}
\arrow[r,"\textrm{\small vary}"] 
\arrow[d,"\textrm{\small symmetrize}"]
& 
\begin{array}{c}\textrm{\small Einstein}\\[-9pt] \textrm{\small equations}\end{array}
\arrow[d,"\textrm{\small symmetrize}"]
& 
\\
\begin{array}{c}\textrm{\small symmetrized}\\[-9pt] \textrm{\small closure}\\[-9pt] \textrm{\small equations}\end{array}
\arrow[r,"\textrm{\small solve}"] 
& 
\begin{array}{c}\textrm{\small symmetrized}\\[-9pt] \textrm{\small Hilbert}\\[-9pt] \textrm{\small action}\end{array}
\arrow[r,"\textrm{\small vary}"] 
& 
\begin{array}{c}\textrm{\small Friedmann}\\[-9pt] \textrm{\small equations}\end{array}
\arrow[r,"\textrm{\small solve}"] 
& 
\begin{array}{c}\textrm{\small FRW}\\[-9pt] \textrm{\small spacetime}\end{array}
\end{tikzcd}
\end{equation*}
\caption{The three paths to arrive at Friedmann's equations by way of gravitational closure of the Klein-Gordon action. The calculationally easiest one (down-down-right-right) is by direct symmetrization of the closure equations and correspondingly simple solution of the latter. The first of two similarly hard ways follows the path (down-right-down-right) and symmetrizes the action which must be obtained previously by general solution of the closure equations. The second hard way (down-right-right-down) symmetrizes the Einstein equations after general solution of the closure equations and variation of the latter.\label{figure}}
\end{figure}
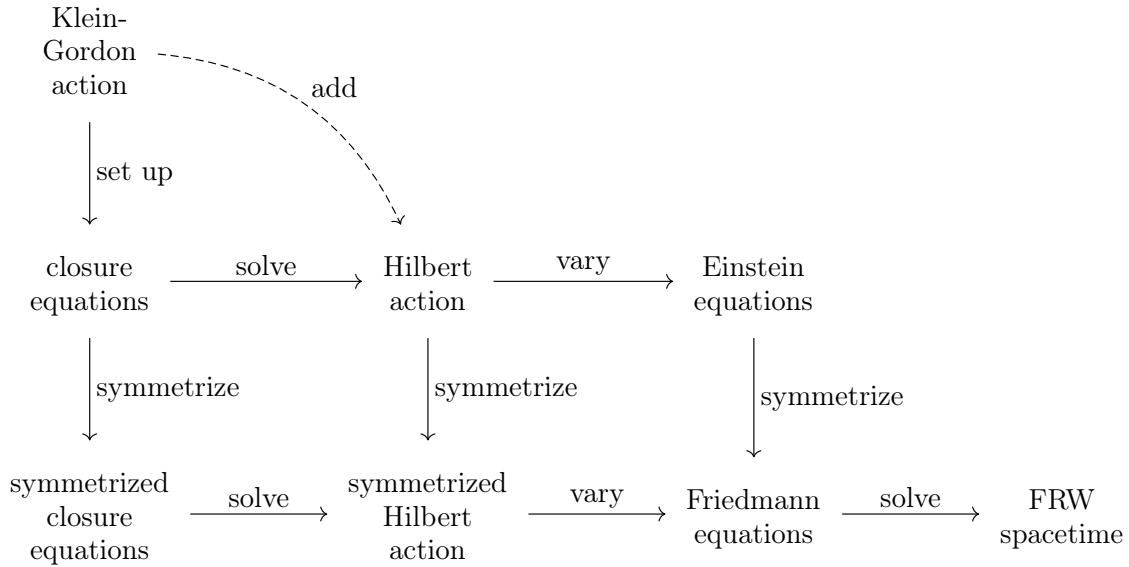

Since the technical steps for the imposition of Killing symmetries on gravitational closure equations do not depend on the actual geometry, \new{essentially because} the Killing condition invariably reads
$$\mathcal{L}_K G = 0$$
for any independent Killing vector field $K$ and any tensorial geometry $G$, we may well illustrate those steps and indeed the practical feasibility of the method for the simplest possible case that allows to perform the symmetrization by explicit calculation at any level (namely that of the gravitational closure equations, the gravitation action or the gravitational field equations, respectively). To this end we choose the \new{otherwise familiar} Klein-Gordon action on a metric geometry as the assumed matter dynamics and impose the maximal symmetry by assuming a spatially homogeneous and isotropic spacetime;  compare Fig. \ref{figure}. This avoids any unnecessary technical overhead,  allows to compare the respective results and yields, as an amusing aside, a sound derivation of Friedman's equations without the need to \new{ever} know Einstein's equations.

The now following technical part of this letter heavily leans on the results of the extensive article \cite{GravClosure18}, but at the same time may serve as a technically particularly accessible introduction to the gravitational closure procedure. We proceed in three steps: First, we prepare to set up the gravitational closure equations for Klein-Gordon theory without any symmetry reduction. Secondly, we weave the symmetry reduction by way of a chain rule into the closure equations and thus reduce the latter. Thirdly, we solve for the symmetry-reduced gravitational Lagrangian and directly obtain the Friedmann equations.

We start from the gravitational closure equations for Klein-Gordon theory on a Lorentzian metric background geometry. This presented the simple example in section V.A of \cite{GravClosure18}. There the six independent geometric configuration fields $\varphi^A=(\varphi^1,\dots,\varphi^6)$ were chosen in order to parametrize the spatial part 
\begin{equation}\label{eq:spatial_frw_metric}
g^{\alpha\beta} = \left[\begin{array}{ccc} \varphi^1  & \varphi^2 & \varphi^3\\
\varphi^2  & \varphi^4 & \varphi^5\\
\varphi^3  & \varphi^5 & \varphi^6
\end{array}\right]^{\alpha\beta}
\end{equation}
of the full inverse spacetime metric, with the latter being completed by the information from  one spatial lapse function $N$ and one spatial shift vector field $\mathcal{N}$. The lapse, the shift and all configuration fields additionally depend on the foliation time. The gravitational closure equations  are the countably infinite set of linear homogeneous partial differential equations displayed in the appendix to \cite{GravClosure18}, which must be solved for functions $C_{A_1\dots A_M}$ that depend locally on the configuration fields~$\varphi^A$. By local dependence we mean dependence on the fields and up to finitely many spatial derivatives of these fields; dependence only on the fields, but not on any derivatives of these, we call ultralocal dependence. The closure equations take the same form for any matter theory and any geometric background, but feature four coefficient functions that capture all relevant information about the matter theory. Since we will need these coefficient functions in their symmetry-reduced form only, we turn to the cosmological symmetry reduction before displaying them here.

Imposing spatial isotropy and homogeneity restricts the lapse to a function of foliation time only and eliminates the shift vector field and all but the three configuration fields
\begin{equation*}
    \varphi^1 = -\frac{1-kr^2}{a^2}\quad,\quad \varphi^4 = -\frac{1}{a^2\,r^2}\quad\mathrm{and}\quad\varphi^6 = - \frac{1}{a^2\,r^2\sin^2\theta}
\end{equation*}
which are given in terms of only one scale factor $a$, which  depends on foliation time only, and spatially polar coordinates $r, \theta, \phi$; their range depends on the only ambiguity left by the symmetry condition, namely on whether the universe is spatially spherical ($k=1$), flat ($k=0$) or hyperbolic ($k=-1$).

How can we evaluate the gravitational closure equations only for such symmetry-reduced configurations? In order to not lose any information, we start with the full set of closure equations for the local functions $C_{A_1\dots A_M}[\varphi^A]$, in the form they take before any symmetry reduction, and then aim to rewrite them in terms of new ultralocal functions
\begin{equation}\label{Ccosmo}
    \CoeffIndex{A_1\dots A_M}^\mathrm{cosmo}(a,r,\theta) := \CoeffIndex{A_1\dots A_M}[\varphi^A(a,r,\theta)]\,,
\end{equation}
where $\varphi^A(a,r,\theta) = (-a^{-2}(1-kr^2),\, 0, \,0,\, -a^{-2}\,r^{-2},\, 0,\, - a^{-2}\,r^{-2}\sin^{-2}\theta )^A$. The reason why the thus defined functions $C^\mathrm{cosmo}_{A_1\dots A_M}$ depend only ultralocally on its independent variables, while the $C_{A_1\dots A_M}$ depended on theirs locally, is simply because homogeneity makes the scale factor $a$ spatially constant. Note that Eq. (\ref{Ccosmo}) does not lend itself to straightforward substitution in all those places where derivatives of the functions $C_{A_1\dots A_M}$ appear in the full closure equations, since these derivatives are taken with respect to the independent configuration fields $\varphi^A$ instead of the independent variables $a, r, \theta$ of $C^{\mathrm{cosmo}}_{A_1\dots A_M}$. The way to relate these derivatives rather requires to employ the chain rule, which provides differential equations for the functions~$\CoeffIndex{A_1\dots A_N}^\mathrm{cosmo}$:
\begin{align}
    \frac{\partial \CoeffIndex{A_1\dots A_N}^\mathrm{cosmo}}{\partial a} &= \left.\Cdiff{A_1\dots A_N}{B}{\mu_1\dots \mu_R}\right|_{\varphi^A(a,r,\theta)}\,\frac{\partial\varphi^B{}_{,\mu_1\dots \mu_R}}{\partial a} \label{eq:pde1_exp_coeff}\,,\\
    \frac{\partial \CoeffIndex{A_1\dots A_N}^\mathrm{cosmo}}{\partial r} &= \left.\Cdiff{A_1\dots A_N}{B}{\mu_1\dots \mu_R}\right|_{\varphi^A(a,r,\theta)}\,\frac{\partial\varphi^B{}_{,\mu_1\dots \mu_R}}{\partial r} \label{eq:pde2_exp_coeff}\,,\\
    \frac{\partial \CoeffIndex{A_1\dots A_N}^\mathrm{cosmo}}{\partial \theta} &= \left.\Cdiff{A_1\dots A_N}{6}{\mu_1\dots \mu_R}\right|_{\varphi^A(a,r,\theta)}\,\frac{\partial\varphi^6{}_{,\mu_1\dots \mu_R}}{\partial a} \,, \label{eq:pde3_exp_coeff}
\end{align}
where the notation $\Cdiff{A_1\dots A_M}{B}{\mu_1\dots\mu_R}$ stands for the partial derivative of the function $C_{A_1\dots A_M}$ with respect to the $R$-fold spatial partial derivative $\partial^R_{\mu_1\dots\mu_R} \varphi^A$ of the configuration field $\varphi^A$ and $|_{\varphi^A(a,r,\theta)}$ denotes evaluation on the symmetric configuration. It remains, before one can solve for the $\CoeffIndex{A_1\dots A_M}^\mathrm{cosmo}$, to symmetry-reduce the coefficient functions $M^{A\gamma}$, $\Ecoeff{A}{\mu}$,  $p^{\mu\nu}$ and $\Fcoeff{A}{\mu}{\gamma}$, which feature in the gravitational closure equations and encode the information conveyed to the latter by the matter theory at hand. Their general form is given in~\cite{GravClosure18}. For our present case, $M^{A\gamma} = 0$, $\Ecoeff{A}{\mu} = \partial_\mu\varphi^A$, $p^{\mu\nu}$ is just the spatial metric $g^{\mu\nu}$ displayed in Eq.~\eqref{eq:spatial_frw_metric} and the fourth coefficient is
\begin{align*}
    \Fcoeff{A}{\mu}{\gamma}|_{\varphi^A(a,r,\theta)} =  &-\left(2\,\delta^A_1\delta^r_\mu\delta^\gamma_r +\sqrt{2}\left(\delta^A_2\delta^\theta_\mu\delta^\gamma_r + \delta^A_3\delta^\varphi_\mu\delta^\gamma_r\right)\right)\,a^{-2}\,(1-kr^2)  \\
    &- \left( 2\, \delta^A_4 \delta^\theta_\mu\delta^\gamma_\theta + \sqrt{2}\,\left(\delta^A_2\delta^r_\mu\delta^\gamma_\theta + \delta^A_5\delta^\varphi_\mu\delta^\gamma_\theta\right)\right)\,a^{-2}\,r^{-2} \\ 
    &-\left(2\,\delta^A_6 \delta^\varphi_\mu\delta^\gamma_\varphi + \sqrt{2}\,\left(\delta^A_3\delta^r_\mu\delta^\gamma_\varphi + \delta^A_5\delta^\theta_\mu\delta^\gamma_\varphi\right)\right)\,a^{-2}\,r^{-2}\sin^{-2}\theta\,
\end{align*}
after cosmological symmetry reduction. Finally, one solves the thus rewritten gravitational closure equations for the functions $\CoeffIndex{A_1\dots A_M}^\mathrm{cosmo}(a,r,\theta)$, which then yields the symmetry-reduced action 
 \begin{equation}\label{eq:ST_action} 
 S_\mathrm{cosmo}(a, \dot a, N) = \int\mathrm dt\int\limits_\Sigma \mathrm d^3z\,\,\sum\limits_{M=0}^\infty C^\textrm{cosmo}_{A_1\dots A_M}(a,r,\theta)\,\dot{\varphi}^{A_1}(a,r,\theta)\cdots\dot{\varphi}^{A_M}(a,r,\theta) N^{1-M}\,,
 \end{equation} 
which is the form to which the action (90) of Ref.~\cite{GravClosure18} simplifies after symmetry reduction. Its variation with respect to the scale factor $a$ and the lapse function $N$ then yield the two Friedmann equations. The functions~$C_{A_1\dots A_N}^\mathrm{cosmo}$ serve as the coefficients of a series expansion in this action and are thus called the expansion coefficients.

Having laid out the strategy to rewrite the gravitational closure equations in terms of the $\CoeffIndex{A_1\dots A_M}^\mathrm{cosmo}$, we now turn to the actual rewriting of all closure equations ($C1$) to ($C7$) 
and ($C8_{N\geq 2}$) to ($C21_{N\geq 2}$)
of Ref.~\cite{GravClosure18} for the case at hand. The first step is to recognize that due to closure equation~$(C8_2)|_{\varphi^A(a,r,\theta)}$, we can express all derivatives $\Cdiff{}{A}{\mu}|_{\varphi^A(a,r,\theta)}$ in terms of six independent derivatives of the type $\Cdiff{}{A}{\mu\nu}|_{\varphi^A(a,r,\theta)}$ which one identifies from the analysis of closure equation~$(C8_3)|_{\varphi^A(a,r,\theta)}$. 
 Besides, combining the derivative $(C4)_{:D}{}^{\lambda_1\lambda_2\lambda_3}\evalphi$ of closure equation~$(C4)$ with the relations obtained from $(C8_3)\evalphi$, we see that the expansion coefficient $C$ depends at most linearly on second derivatives of the configuration fields --- when evaluated on the symmetric configuration. As a direct consequence of this and the aforementioned closure equation $(C8_2)\evalphi$, the expansion coefficient $C$ also contains no terms with both first and second derivatives of the $\varphi^A$.

A further consequence of this and closure equation~$(C3)$ is that the expansion coefficient $\CoeffIndex{AB}\evalphi$ depends only on the configuration fields, but not on any spatial derivatives. This is the key towards a solution for the expansion coefficient~$\CoeffIndex{AB}^\mathrm{cosmo}$ as the corresponding differential equations~\eqref{eq:pde1_exp_coeff} --- \eqref{eq:pde3_exp_coeff} for the component $\CoeffIndex{14}^\mathrm{cosmo}$ reduce to
\begin{align*}
    \frac{\partial \CoeffIndex{14}^\mathrm{cosmo}}{\partial a} &= \frac{2(1-kr^2)}{a^3}\,\Cdiff{14}{1}{}\evalphi + \frac{2}{a^3\,r^2}\,\Cdiff{14}{4}{}\evalphi + \frac{2}{a^3\,r^2\sin^2\theta}\,\Cdiff{14}{6}{}\evalphi \,, \\
    \frac{\partial \CoeffIndex{14}^\mathrm{cosmo}}{\partial r} &= \frac{2\,kr}{a^2}\,\Cdiff{14}{1}{}\evalphi + \frac{2}{a^2\,r^3}\,\Cdiff{14}{4}{}\evalphi + \frac{2}{a^2\,r^3\sin^2\theta}\,\Cdiff{14}{6}{}\evalphi \,, \\
    \frac{\partial \CoeffIndex{14}^\mathrm{cosmo}}{\partial\theta} &= \frac{2\,\cos\theta}{a^2\,r^2\sin^3\theta}\,\Cdiff{14}{6}{}\evalphi \,,
\end{align*}
which can be solved using relations from~$(C10_2)\evalphi$. We get
\begin{equation*}
    \CoeffIndex{14}^\mathrm{cosmo} = K_0\,\frac{r^4\sin\theta\,a^7}{(1-kr^2)^\frac{3}{2}}
\end{equation*}
with one constant of integration $K_0$. We can calculate the other two relevant non-trivial components $\CoeffIndex{16}^\mathrm{cosmo}$ using the solution of $\CoeffIndex{14}^\mathrm{cosmo}$ to be
\begin{equation*}
    \CoeffIndex{16}^\mathrm{cosmo} = K_0\,\frac{r^4\,\sin^3\theta\,a^7}{(1-kr^2)^\frac{3}{2}} \quad\mathrm{and}\quad\CoeffIndex{46}^\mathrm{cosmo} = K_0\,\frac{r^6\sin^3\theta\,a^7}{(1-kr^2)^\frac{1}{2}}\,.
\end{equation*}
All other components of the expansion coefficient $\CoeffIndex{AB}^\mathrm{cosmo}$ either vanish or are not relevant for the Lagrangian as they couple to vanishing $\dot{\varphi}^A$ in the expansion~\eqref{eq:ST_action}.

We now use the solution of $\CoeffIndex{AB}^\mathrm{cosmo}$ and determine the derivatives $\Cdiff{}{A}{\mu\nu}\evalphi$ using closure equation~$(C3)\evalphi$. Subsequently, we insert the resulting expressions first into $(C8_2)\evalphi$ and then $(C1)|_{\varphi^A(a,r,\theta)}$ in order to express all terms of the differential equations for $C^\mathrm{cosmo}$. In the end, we are left with three differential equations from~\eqref{eq:pde1_exp_coeff} --- \eqref{eq:pde3_exp_coeff}; their solution amounts to the coefficient
\begin{equation*}
    C^\mathrm{cosmo} = \frac{r^2\sin\theta}{(1-kr^2)^\frac{1}{2}}\,\left(K_1\,a^3 - 24\,K_0\,ka\right)
\end{equation*}
with a second constant of integration $K_1$.

After the determination of these two expansion coefficients --- it will turn out that they are the only relevant ones --- we turn towards the sequence $(C16_{N\geq 2})$ of equations. Evaluating all instances of this sequence with even $N$ shows that all odd-numbered expansion coefficients $C_{A_1\dots A_{2M+1}}^\mathrm{cosmo}$ for $M\geq 1$ vanish. The same analysis of all instances with odd $N$ reveals that also all even-numbered expansion coefficients $C_{A_1\dots A_{2M}}^\mathrm{cosmo}$ for $M \geq 2$ vanish. This leaves us with a separate set of equations for the last remaining expansion coefficient $C_A^\mathrm{cosmo}$ which just is a boundary term that can be dropped from the Lagrangian --- similar to the solution of the closure equations from Maxwell theory without a symmetry assumption \cite{PractitionersGuide, DuellPhD}. We can now construct the spacetime action by inserting the two coefficients $C^\mathrm{cosmo}$ and $C_{AB}^\mathrm{cosmo}$ into the expansion~\eqref{eq:ST_action}.

The above solution of the gravitational closure equations has given us two constants of integration which need to be determined experimentally. At the level of the Friedmann equations, these two constants will be identified as Newton's constant $G$ and the cosmological constant $\Lambda$. Using this identification, the cosmological action~$S_\mathrm{cosmo}$ constructed from the solution of the symmetry-reduced closure equations is
\begin{equation*}
    S_\mathrm{cosmo} = \int\mathrm dt\int\limits_\Sigma \mathrm d^3 x\, \frac{r^2\sin\theta}{(1-kr^2)^\frac{1}{2}}\left[N\,\left(\frac{\Lambda\,a^3}{8\pi\,G} - \frac{3\,k a}{8\pi\,G}\right) + \frac{3}{8\pi\,G}\,\frac{a\dot{a}^2}{N}\right]\,,
\end{equation*}
which is to be varied with respect to the two symmetric geometric degrees of freedom, the scale factor $a(t)$ and the lapse function $N(t)$.

\new{The gravitational dynamics are to be sourced by the expression on the right hand side of the second equation in (\ref{alleqns}), which in the present case is just one half times the familiar energy momentum tensor density
\begin{equation*}
    \widetilde T_{ab} := - 2\,\frac{\delta S_\mathrm{matter}}{\delta g^{ab}}\,.
\end{equation*}
As is well known, however, the assumption of cosmological symmetries is meant to hold at large scales only, since their imposition at all scales might not even yield any non-trivial right hand side of the second equation of (\ref{alleqns}). Instead, an effective `perfect fluid' energy-momentum
tensor density --- which for the Lorentzian metric geometry of our present calculation takes the form 
$$\widetilde T^\textrm{\tiny eff}_{ab} = \sqrt{-g}\big((\rho+p) u_a u_b + pg_{ab}\big)\,$$
 in order to be compatible with spatial homogeneity and isotropy for spatially constant $\rho$ and $p$ --- must be obtained by appropriate averaging over matter field configurations. The small-scale matter theory then merely determines an equation of state that relates $\rho$ and $p$.
This standard reasoning directly generalizes \cite{Fischer2017} {\it mutatis mutandis} to general tensorial spacetime geometries $G$, which typically require further spatially constant functions beyond $\rho$ and $p$.
Including such effective perfect fluid matter,} the two Friedmann equations arising from our solution of the symmetry-reduced gravitational closure equations for Klein-Gordon theory are
\begin{equation*}
    \left(\frac{\dot{a}}{a}\right)^2 = \frac{8\pi\,G}{3}\,\rho - \frac{k}{a^2} + \frac{\Lambda}{3}\quad\mathrm{and}\quad \frac{\ddot{a}}{a} = - \frac{4\pi\,G}{3}(\rho + 3\,p) + \frac{\Lambda}{3}\,.
\end{equation*}
We stress here again, since this is the point of the present letter, that the derivation of the Friedmann equations never involved the Einstein equations or their symmetry reduction. Instead, we set up the gravitational closure equations for Klein-Gordon dynamics and solved those for geometric configurations that correspond to spatially isotropic and homogeneous spacetime metrics.

The method to obtain a symmetry reduction of the gravitational closure equations, which we illustrated above for the special case of a Klein-Gordon action on a metric background as the input matter dynamics, generalizes immediately to gravitational closure equations where the input matter model reaches beyond the current standard model of particle physics, while gravitational closure of the latter reassuringly leads to standard general relativity \cite{Wierzba}.
The initially mentioned general linear electrodynamics, built on a geometry described by a fourth-rank tensor field, is a physically interesting example for non-standard model matter in search for the underlying gravitational theory, since the latter is needed in order to quantitatively predict where birefringence effects occur. The corresponding coefficient functions for the gravitational closure equations are quickly calculated, but sufficiently involved to push a general solution to the resulting closure equations out of immediate reach. Symmetry reductions of the pertinent closure equations, however, simplify the equations to a point where solving the them by hand becomes more realistic, commensurate with the degree of symmetry. Cosmological dynamics will thus always be the easiest ones to obtain by gravitational closure of any given matter model.

Certainly, the most interesting application of gravitational closure, with or without symmetry assumptions, will of course arise if and when the standard model of particle physics will require a phenomenologically inescapable extension that employs a geometric background more refined than the one provided by a Lorentzian metric geometry. In fact, this may be phenomenologically less exotic than one might assume at first sight, see \cite{GST}. The question of what the underlying gravity theory will be then reduces to the problem of solving the corresponding gravitational closure equation. The symmetry reduction presented here will then provide one of the main tools to obtain exact gravitational field equations.

\section*{Acknowledgements}
The authors thank Marcus Werner, Florian Wolz and Alexander Wierzba for most valuable comments and discussions. MD gratefully acknowledges financial support by the Graduate Academy of Heidelberg University and the Heidelberg Graduate School for Physics. FPS and NF would like to thank Marcus Werner and the Yukawa Institute for the hospitality and financial support during repeated invitations to extensive research stays in Kyoto. NF further acknowledges support from the DAAD PROMOS program for this visit.

\bibliography{references}{}
\end{document}